\documentclass[journal]{IEEEtran}

\usepackage{graphicx}  

\usepackage{amsmath}   
\usepackage{amsfonts}

\usepackage{latexsym}

\newcommand{\mean}[1]{\langle{#1}\rangle}

\newcommand{\bra}[1]{\langle{#1}|}
\newcommand{\ket}[1]{|{#1}\rangle}

\newcommand{\dgg}{^{\dagger}}

\begin{document}

\title{
Experimental demonstration of coherent feedback control 
on optical field squeezing
}

%


\author{Sanae Iida, Mitsuyoshi Yukawa, Hidehiro Yonezawa, 
        Naoki Yamamoto, and Akira Furusawa
\thanks{S. Iida, M. Yukawa, H. Yonezawa, and A. Furusawa are with the 
        Department of Applied Physics, School of Engineering, 
        The University of Tokyo, 7-3-1, Hongo, Bunkyo-ku, Tokyo 113-8656, 
        Japan       
        (e-mail: sanae111@gmail.com, yukawa@alice.t.u-tokyo.ac.jp, 
        yonezawa@ap.t.u-tokyo.ac.jp, akiraf@ap.t.u-tokyo.ac.jp). 
        S. Iida and N. Yamamoto are with the Department of 
        Applied Physics and Physico-Informatics, Keio University, 
        Hiyoshi 3-14-1, Kohoku-ku, Yokohama 223-8522, Japan 
        (e-mail: yamamoto@appi.keio.ac.jp). }
\thanks{
This work was partly supported by PDIS, GIA, G-COE, APSA, JSPS, 
FIRST commissioned by the MEXT of Japan, and SCOPE program of the 
MIC of Japan. 
}
}

\maketitle

%
%
%
%
%
%
%


\vspace{0.7cm}

\begin{abstract}



Coherent feedback is a non-measurement based, hence a back-action 
free, method of control for quantum systems. 
A typical application of this control scheme is squeezing enhancement, 
a purely non-classical effect in quantum optics. 
In this paper we report its first experimental demonstration that 
well agrees with the theory taking into account time delays and 
losses in the coherent feedback loop. 
The results clarify both the benefit and the limitation of coherent 
feedback control in a practical situation.

\end{abstract}


%

\IEEEpeerreviewmaketitle

\section{Introduction}

Feedback control theory has recently been further extended to cover 
even quantum systems such as a single atom. 
The methodologies are broadly divided into two categories: 
measurement-based feedback control and {\it non}-measurement-based 
one called the {\it coherent feedback control}. 
Below we briefly describe their major difference and particularly 
a specific feature of the latter control strategy.

Quantum (continuous-time) measurement produces useful information 
(which is of course a continuous-time signal) that can be fed back 
to the system of interest, though at the same time it introduces 
unavoidable {\it back-action} noise into the system 
\cite{Belavkin-1992,Bouten-2007}. 
A number of investigation of this trade-off have discovered several 
situations where the measurement-based feedback control has clear 
benefits \cite{Wiseman-2009}, for instance an application to 
quantum error correction \cite{Ahn-2002}. 
On the other hand, the coherent feedback (CF) control 
\cite{Gough-2009b,Gough-2010,Lloyd-2000,Wiseman-1994b,Yanagisawa-2003a,
Yanagisawa-2003b} takes a totally different approach. 
The general structure of the CF control is shown in Fig. 1 (a); 
the system outputs a ``quantum signal", then the controller, which 
is also a quantum system, coherently modulates the output and 
feeds it back to control the system. 
In this scheme any measurement is not performed, implying that 
no excess measurement back-action noise is introduced into the 
system. 
Because of this feature the CF control is suitable for dealing 
with problems of {\it noise reduction}, which is the central topic 
in the control theory. 
Actually we find that the very successful noise-reducing controllers, 
the $H^\infty$ and the Linear Quadratic Gaussian controllers, 
have natural CF control analogues \cite{James-2008,Maalouf-2009,
Maalouf-2010,Nurdin-2009,Yanagisawa-2003b}.

\begin{figure}[h]
\centering
\includegraphics[width=1.0\linewidth]{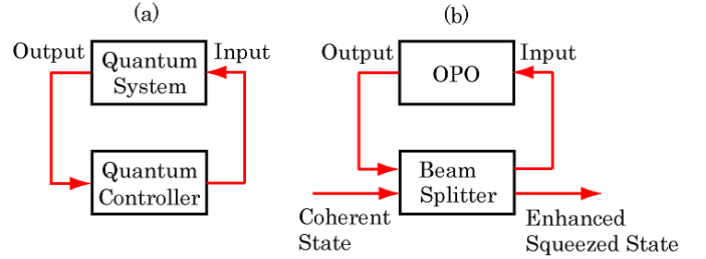}
\caption{
(a) General structure of the CF control. 
The arrows correspond to unidirectional flows of ``quantum signals" 
such as optical laser fields. 
(b) Optical system structure of the CF control for squeezing 
enhancement, corresponding to Fig. 3.}
\label{fig:abstractCF}
\end{figure}

Here we mention about a {\it squeezed state} \cite{Walls-1983}; 
this is a purely non-classical state where the fundamental quantum 
noise is reduced below the quantum noise limit in one of the 
quadrature observables such as the position $\hat{q}$ and the 
momentum $\hat{p}$ (a more detailed description will be given in 
Section II). 
Therefore, a situation in which noise reduction via the CF control 
shows the best efficiency may appear in problems of generating 
squeezed states. 
Actually Yanagisawa \cite{Yanagisawa-2003b} and Gough 
\cite{Gough-2009} theoretically showed that this idea works in 
the case of quantum optics. 
Fig.~1~(b) illustrates the CF loop structure they studied. 
The quantum system, which now corresponds to an {\it optical 
parametric oscillator (OPO)}, has an ability of noise reduction; 
that is, it transforms an input {\it coherent state} into an 
output squeezed state. 
It was then shown that the performance of squeezing can be 
enhanced by constructing an appropriate CF controller, 
which in this case is given by a {\it beam splitter} with tunable 
transmissivity.

The purpose of this paper is to report the first experimental 
demonstration of the above-mentioned CF control on optical field 
squeezing, which well agrees with the theory that carefully takes 
into account the effects of the actual laboratory setup, 
particularly time delays and losses in the feedback loop. 
The results are significant in the sense that, both theoretically 
and experimentally, they clarify the situation where the CF control 
is really effective and the limitation on how much it can 
improve the system performance practically. 
Note that in \cite{Gough-2009,Yanagisawa-2003b} a realistic closed 
loop model was not considered, hence such benefit and limitation 
are first clarified in this paper. 
Another important remark is that, while Mabuchi has experimentally 
demonstrated classical noise reduction with the CF control 
\cite{Mabuchi-2008}, in our case we deal with purely non-classical 
noise reduction that beats the quantum noise limit, which is crucial 
in quantum information processing. 
In this sense, this paper provides the first experimental 
demonstration of realistic applicability of the CF control 
to non-classical regime.


\section{Preliminaries}

In this section we provide some notions of quantum mechanics 
and the dynamics of an OPO. 
For more details see \cite{Bachor-2004,Collet-1984,Furusawa-2011,
Gardiner-1984,Gardiner-2004,Leonhardt-2010}.

\subsection{Observables, states, and statistics in quantum optics}

In quantum mechanics, unlike the classical case, physical quantities 
must take different values probabilistically when measuring them. 
The corresponding statistics is described in terms of a {\it state}, 
which is represented by a unit vector $\ket{\psi}$. 
In general, when measuring a physical quantity represented by a 
self-adjoint operator $\hat{X}=\hat{X}\dgg$, the mean and variance 
of the measurement results are respectively given by 
\begin{equation*}
\label{mean variance def}
    \mean{\hat{X}}:=\bra{\psi}\hat{X}\ket{\psi},~~~
    \mean{\Delta\hat{X}^2}
         :=\bra{\psi}\hat{X}^2\ket{\psi}
             -\bra{\psi}\hat{X}\ket{\psi}^2, 
\end{equation*}
where $\bra{\psi}$ is the adjoint to $\ket{\psi}$. 
That is, a state corresponds to a probability distribution. 
For these statistical values to take real numbers, $\hat X$ must 
be self-adjoint; 
actually in quantum mechanics any physical quantity is represented 
by a self-adjoint operator and is called an {\it observable}.

A single-mode quantum optical field is described with an operator 
$\hat a$ and its adjoint $\hat a^\dagger$, which respectively 
correspond to a complex amplitude and its conjugate of a classical 
optical field. 
These operators satisfy the {\it canonical commutation relation (CCR)} 
$[\hat{a},\hat{a}\dgg]=\hat{a}\hat{a}\dgg-\hat{a}\dgg\hat{a}=1$. 
These field operators are not self-adjoint, i.e., not observables, 
hence let us define 
\begin{equation*}
\label{quadratures}
    \hat{x}_{+}:=\hat{a}+\hat{a}\dgg,~~~
    \hat{x}_{-}:=-i(\hat{a}-\hat{a}\dgg). 
\end{equation*}
Analogous to the classical case, these observables are called 
the {\it amplitude} and {\it phase quadratures}. 
The CCR for these observables is $[\hat{x}_{+}, \hat{x}_{-}]=2i$; 
due to this equality, for any state the variances satisfy the 
{\it Heisenberg's uncertainty relation}: 
\begin{equation}
\label{uncertainty relation}
   \mean{\Delta\hat{x}_{+}^2}\mean{\Delta\hat{x}_{-}^2}\geq 1. 
\end{equation}
This means that the amplitude and phase quadratures cannot be 
determined simultaneously. 
In other words, there is fundamental uncertainty with respect to 
these two non-commutative observables. 
In particular, it is known that, for any {\it classical} state 
such as a thermal state, each variance must be bigger than $1$, i.e., 
$\mean{\Delta\hat{x}_{+}^2}\geq 1$ and 
$\mean{\Delta\hat{x}_{-}^2}\geq 1$; 
this lower bound is called the {\it quantum noise limit (QNL)}. 
Related to this limit, we here introduce two important states in 
quantum optics: coherent and squeezed states. 
A coherent state is the closest possible analogue to a classical 
electromagnetic wave, which is generated with a laser device. 
The coherent state $\ket{\alpha}$ is defined as an eigenvector of 
${\hat a}$ with $\alpha\in{\bf C}$ the corresponding eigenvalue, 
i.e., ${\hat a}\ket{\alpha}=\alpha\ket{\alpha}$. 
Note that $\ket{0}$ represents a vacuum state.
A crucial property of $\ket{\alpha}$ is that it achieves the QNL, 
i.e., we have 
$\mean{\Delta\hat{x}_{+}^2}=\mean{\Delta\hat{x}_{-}^2}=1$ 
for any $\alpha$, meaning that a coherent state is a lowest-noise 
classical state. 
On the other hand, a squeezed state is a purely non-classical state 
with one of the quadrature variance below the QNL. 
In particular, for an ideal pure squeezed state the quadrature 
variances are given by $\mean{\Delta\hat{x}_{+}^2}=e^{2r}$ and 
$\mean{\Delta\hat{x}_{-}^2}=e^{-2r}$ with $r\in{\bf R}$ a unit-less 
parameter. 
Squeezed states play important roles in quantum information 
technologies, because for instance {\it entanglement}, which is a 
key property to perform various quantum information processing, 
can be generated using squeezed states. 
Quantum information processing often relies on highly entangled 
states, and this equivalently means highly squeezed states are 
desirable. 
This is the reason why squeezing enhancement is of vital important.


\subsection{Optical parametric oscillator as a linear system}

\begin{figure}[h]
  \centering
  \includegraphics[width=0.9\linewidth]{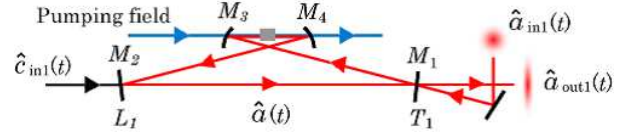}
  \caption{Schematic of the OPO. Arrows represent optical 
  beams travelling along those directions. $M_i$ are mirrors 
  $(i=1,\cdots, 4)$. 
  A crystal between the curved mirrors $M_3$ and $M_4$ is 
  a second-order nonlinear medium. 
  }
  \label{OPO}
\end{figure}

The {\it optical parametric process} is a widely-used method for 
generating a squeezed optical field, where a coherent optical field 
is squeezed by the interaction with a strong pumping field through 
a nonlinear medium. 
To produce a squeezed state of light more effectively, it is common 
to use the {\it optical parametric oscillator (OPO)}; a cavity that 
contains a second-order nonlinear crystal. 
Fig. \ref{OPO} shows a conventional bow-tie type cavity with 
four mirrors, where the mirror $M_1$ (input-output port) is 
partially transmissive and the others are highly reflective for 
the fundamental optical field. 
All the mirrors are perfectly transparent for the pumping field. 
This system renders the input coherent field $\hat{a}_{{\rm in}1}(t)$ 
interact with the medium many times inside the cavity and finally 
produces a well-squeezed field $\hat{a}_{{\rm out}1}(t)$ 
at the output port. 
These input and output fields couple to the internal cavity field 
$\hat{a}(t)$ through the mirror $M_1$ characterized by the power 
transmissivity $T_1$. 
Losses inside the OPO are modeled as a coupling between $\hat a (t)$ 
and an unwanted vacuum field $\hat{c}_{{\rm in}1}(t)$ through one 
of the end mirror $M_2$ with the transmissivity $L_{1}$. 
Here we assume that such interactions occur instantaneously, 
implying that the outer fields satisfy the CCR 
$[\hat{a}_{{\rm in}1}(t),\hat{a}\dgg_{{\rm in}1}(t')]=\delta (t-t')$ and 
$[\hat{c}_{{\rm in}1}(t),\hat{c}\dgg_{{\rm in}1}(t')]=\delta (t-t')$.

Let us describe the dynamics of the cavity field $\hat{a}(t)$, 
which is on resonant with the outer fields. 
In quantum mechanics, any observable ${\hat X}$ changes in time 
according to the {\it Heisenberg equation} 
$d{\hat X}/dt=i[{\hat X}, {\hat H}]$, where $\hat{H}=\hat{H}\dgg$ 
is called the {\it Hamiltonian}. 
Now the Hamiltonian only for $\hat{a}(t)$ is given by 
\begin{equation*}
   \hat{H}=
      \omega_0 \hat{a}\dgg(t)\hat{a}(t)
         +\frac{i}{2}\big[ 
             \epsilon e^{-i 2\omega_0 t}\hat{a}\dgg(t)\mbox{}^2
               -\epsilon^{\ast} e^{i 2\omega_0 t}\hat{a}(t)^2 \big], 
\end{equation*}
where $\omega_0$ is the resonant frequency and $\epsilon$ denotes 
the effectiveness of the nonlinear medium which depends on the 
pumping field strength of frequency $2\omega_0$. 
The Heisenberg equation of $\hat{a}(t)$ that involves the 
coupling to the outer fields is given by the following 
{\it quantum Langevin equation} \cite{Bachor-2004,Collet-1984,
Gardiner-1984,Gardiner-2004}: 
\begin{eqnarray}
& & \hspace*{-1em}
\label{eq-quantum Langevin equation}
   \frac{d \hat{a}(t)}{dt}
     =-i \omega_{0}\hat{a}(t) 
        + \epsilon e^{-2i\omega_{0}t}\hat{a}\dgg(t)
      - \frac{\gamma}{2} \hat{a}(t) 
\nonumber \\ & & \hspace*{5em}
   \mbox{}
   + \sqrt{\gamma_1}\hat{a}_{{\rm in}1}(t) 
      + \sqrt{\gamma_{L1}}\hat{c}_{{\rm in}1}(t), 
\end{eqnarray}
where $\gamma:=\gamma_1+\gamma_{L1}$, and $\gamma_1:=cT_1/l$ and 
$\gamma_{L1}:=cL_{1}/l$ represent the damping rates with $l$ the 
optical path length in the OPO and $c$ the speed of light. 
The outer fields satisfy the following boundary condition:
\begin{equation}
\label{eq-boundary conditions}
    \hat{a}_{{\rm out}1}(t) 
       = \sqrt{\gamma_1}\hat{a}(t) - \hat{a}_{{\rm in}1}(t). 
\end{equation}
The single-input and single-output linear system given by Eqs. 
\eqref{eq-quantum Langevin equation} and 
\eqref{eq-boundary conditions} is the system 
generating a squeezed state of light.

As in the classical case, a simple input-output relation is found 
in the Fourier domain 
$\hat{O}(\Omega)=\int dt \hat{O}(t)e^{i\Omega t}/\sqrt{2\pi}$, 
where we have moved to the rotating frame at frequency $\omega_0$ by 
setting $\hat{O}(t)=\hat{o}(t)e^{i\omega_0 t}$. 
We will deal with for instance $\hat{A}_{{\rm in}1}(\Omega)$ and 
$\hat{A}\dgg_{{\rm out}1}(\Omega)$, that corresponds to 
$\hat{a}_{{\rm in}1}(t)$ and $\hat{a}\dgg_{{\rm out}1}(t)$, 
respectively. 
As a result we have 
\begin{eqnarray}
& & \hspace*{-1em}
\label{A_out(Omega)}
   \hat{A}_{{\rm out}1}(\Omega) = 
      G(\Omega)\hat{A}_{{\rm in}1}(\Omega) 
       + g(\Omega)\hat{A}_{{\rm in}1}\dgg(\Omega)
\nonumber \\ & & \hspace*{5em}
   \mbox{}
  + \bar{G}(\Omega) \hat{C}_{{\rm in}1}(\Omega) 
       + \bar{g}(\Omega)\hat{C}_{{\rm in}1}\dgg(\Omega), 
\end{eqnarray}
where
\begin{align}
   G(\Omega) =& 
     \frac{ (\gamma_1/2)^2 - (\gamma_{L1}/2-i\Omega)^2 + |\epsilon|^2 }
          { (\gamma/2-i\Omega)^2-|\epsilon|^2 },~~
\nonumber
\\
   \bar{G}(\Omega) =& 
     \frac{ \sqrt{\gamma_1 \gamma_{L1}}(\gamma/2-i\Omega) }
          { (\gamma/2-i\Omega)^2 - |\epsilon|^2 },~~
   g(\Omega) = 
     \frac{ \epsilon \gamma_1 }
          { (\gamma/2-i\Omega)^2 - |\epsilon|^2 },
\nonumber
\end{align}
and $\bar{g}(\Omega)=\sqrt{\gamma_{L1}/\gamma_1}g(\Omega)$. 
To evaluate the squeezing, let us introduce the (generalized) 
quadrature in the Fourier domain: 
\begin{equation}
\label{general quadrature}
   \hat{X}^{\theta}_{{\rm out}1}(\Omega) = 
     \frac{1}{2}\big[ e^{i\theta}\hat{A}_{{\rm out}1}(\Omega) 
                    + e^{-i\theta}\hat{A}_{{\rm out}1}\dgg(\Omega) \big].
\end{equation}
We write 
$\hat{X}^{+}_{{\rm out}1}(\Omega)=\hat{X}^0_{{\rm out}1}(\Omega)$ and 
$\hat{X}^{-}_{{\rm out}1}(\Omega)=\hat{X}^{\pi/2}_{{\rm out}1}(\Omega)$. 
For $\hat{A}_{{\rm in}1} (\Omega)$ and $\hat{C}_{{\rm in}1} (\Omega)$ 
their quadratures are defined in the same form as Eq. 
\eqref{general quadrature}. 
Then, from Eq. \eqref{A_out(Omega)} we have 
\begin{equation*}
   \hat{X}_{{\rm out}1}^{\pm}(\Omega) = 
      [G(\Omega)\pm g(\Omega)]\hat{X}^{\pm}_{{\rm in}1}(\Omega)
     +[\bar{G}(\Omega)\pm \bar{g}(\Omega)]\hat{X}^{\pm}_{L1}(\Omega). 
\end{equation*}
The variance of $\hat{X}_{{\rm out}1}^{\pm}(\Omega)$ is simply 
given by the power spectrum 
$S^{\pm}_{{\rm out}1}(\Omega)
:=\langle|\hat{X}^{\pm}_{{\rm out}1}(\Omega)|^2\rangle$. 
In particular, when the input field is a vacuum state, 
we have 
\begin{equation*}
\label{S(Omega)0}
    S^{\pm}_{{\rm out}1}(\Omega) = 
        |G(\Omega)\pm g(\Omega)|^2
           + |\bar{G}(\Omega) \pm \bar{g}(\Omega)|^2. 
\end{equation*}
If $\epsilon=0$, then $S^{\pm}_{{\rm out}1}(\Omega)=1,~\forall\Omega$, 
which is the QNL. 
But a squeezed state of light is generated when $\epsilon\neq 0$; 
actually for simplicity in the case 
$\gamma_{L1}=0, \epsilon\in{\bf R}$, and $\Omega=0$, we have 
\begin{equation*}
    S^{+}_{{\rm out}1}(0) 
      = \Big( \frac{\gamma_1+2\epsilon}{\gamma_1-2\epsilon} \Big)^2,~~~
    S^{-}_{{\rm out}1}(0) 
      = \Big( \frac{\gamma_1-2\epsilon}{\gamma_1+2\epsilon} \Big)^2, 
\end{equation*}
one of which is below the QNL. 
Note that the sign of $\epsilon$ determines which quadrature 
is squeezed or {\it anti}-squeezed.


\section{Coherent feedback control on optical field squeezing}

\begin{figure}[h]
  \centering
  \includegraphics[width=0.8\linewidth]{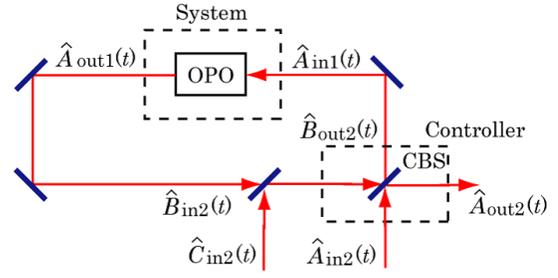}
  \caption{
  Schematic of the CF control on optical field squeezing. 
  }
  \label{fig:CF}
\end{figure}

The basic idea of the CF control for optical squeezing 
enhancement is found in \cite{Gough-2009,Yanagisawa-2003b}. 
We here study a realistic model corresponding to an actually 
constructed optical system in the laboratory, which takes into 
account time delays and losses in the feedback loop.

The CF structure is depicted in Fig. \ref{fig:CF}. 
The system is the OPO described in Sec. II. B. 
In this CF control scheme, a beam splitter (BS) plays the roles 
of both a controller and an input-output port. 
Hereafter we name this BS as the {\it control-BS} (CBS) to 
discern it from the other BSs. 
The transmissivity $T_2$ of the CBS is tuned to obtain higher 
squeezing level.
The coherent input field $\hat{A}_{{\rm in}2}(t)$ is sent to 
one port of the CBS, and then, one of its outputs 
$\hat{B}_{{\rm out}2}(t)$ is sent to the OPO. 
The output of the OPO, $\hat{A}_{{\rm out}1}(t)$, is sent back 
to the CBS to close the loop. 
Finally at the other output port of the CBS we will find an 
enhanced squeezed field $\hat{A}_{{\rm out}2}(t)$. 
The input-output relation at the CBS is given by (in the 
rotating frame)
\begin{align}
    \hat{A}_{{\rm out}2}(t) =& 
        \sqrt{1-T_2}\hat{A}_{{\rm in}2}(t)
\nonumber 
\\ 
        & +\sqrt{T_2} \big[ \sqrt{1-L_{2}} \hat{B}_{{\rm in}2}(t)
          +\sqrt{L_{2}}\hat{C}_{{\rm in}2}(t) \big],
\nonumber 
\\ 
    \hat{B}_{{\rm out}2}(t) =& 
        -\sqrt{1-T_2} \big[ \sqrt{1-L_{2}} \hat{B}_{{\rm in}2}(t)
           +\sqrt{L_{2}}\hat{C}_{{\rm in}2}(t) \big]
\nonumber 
\\ 
        &+\sqrt{T_2} \hat{A}_{{\rm in}2}(t), 
\nonumber
\end{align}
where $\hat{C}_{{\rm in}2}(t)$ is a vacuum field entering through 
a fictitious BS with reflectivity $L_{2}$, and this is a model 
of losses in the CF loop. 
$\hat{B}_{{\rm in}2}(t)$ is the output of the OPO just before 
entering this fictitious BS. 
Here it is assumed that the fictitious BS is placed just before the CBS. 
Now let $\tau_a:={l_a}/{c}$ ($\tau_b:={l_b}/{c}$) be the time delay 
resulting from the optical path length $l_a$ ($l_b$) from (to) 
the CBS to (from) the OPO. 
Then we have 
\begin{equation*}
   \hat{A}_{{\rm in}1}(t) 
        = \hat{B}_{{\rm out}2}(t-\tau_a)e^{i\omega_0 \tau_a},~~
   \hat{B}_{{\rm in}2}(t) 
        = \hat{A}_{{\rm out}1}(t-\tau_b)e^{i\omega_0 \tau_b}. 
\end{equation*}
Combining these equations with Eq. \eqref{A_out(Omega)} the final 
input-output relation is given in terms of the quadrature 
representation by
\begin{eqnarray}
& & \hspace*{-1.9em}
    \hat{X}^{\pm}_{{\rm out}2}(\Omega)
\nonumber \\ & & \hspace*{-1.5em} \mbox{}
    =\Big{[} \sqrt{1-T_2}
      + \frac{T_2\sqrt{1-L_{2}}\alpha^{\pm}(\Omega)}
            {1+\alpha^{\pm}(\Omega) \sqrt{(1-T_2)(1-L_{2})} } \Big{]}
               \hat{X}^{\pm}_{{\rm in}2}(\Omega)
\nonumber \\ & & \hspace*{-1.2em} \mbox{}
    + \frac{\sqrt{T_2(1-L_{2})}\beta^{\pm}(\Omega)}
           {1+\alpha^{\pm}(\Omega) \sqrt{(1-T_2)(1-L_{2})}}
             \hat{X}^{\pm}_{L1}(\Omega) 
\nonumber \\ & & \hspace*{-1.2em} \mbox{}
    + \Big{[} \sqrt{T_2 L_{2}}
    - \frac{\sqrt{T_2(1-L_{2})(1-T_2)L_{2}}\alpha^{\pm}(\Omega)}
       {1+\alpha^{\pm}(\Omega) \sqrt{(1-T_2)(1-L_{2})} } \Big{]}
 \hat{X}^{\pm}_{L2}(\Omega), 
\nonumber
\end{eqnarray}
where 
$\alpha^{\pm}(\Omega) = [G(\Omega)\pm g(\Omega)]
e^{i(\Omega+\omega_0)(\tau_a+\tau_b)}$ and 
$\beta^{\pm}(\Omega)  = [\bar{G}(\Omega) \pm \bar{g}(\Omega)]
e^{i(\Omega+\omega_0)\tau_b}$. 
When the input is a vacuum state, the power spectrum 
$S^{\pm}_{{\rm out}2}(\Omega)
:=\langle|\hat{X}^{\pm}_{{\rm out}2}(\Omega)|^2\rangle$ 
is given by
\begin{eqnarray}
& & \hspace*{-2em}
   S^{\pm}_{{\rm out}2}(\Omega) = 
     \Big| 
        \sqrt{1-T_2} 
        + \frac{T_2\sqrt{1-L_{2}}\alpha^{\pm}(\Omega)}
               {1+\alpha^{\pm}(\Omega) \sqrt{(1-T_2)(1-L_{2})}} \Big|^2
\nonumber \\ & & \hspace*{2.5em} \mbox{}
    + \frac{T_2(1-L_{2})|\beta^{\pm}(\Omega)|^2}
          {|1+\alpha^{\pm}(\Omega) \sqrt{(1-T_2)(1-L_{2})}|^2}
\nonumber \\ & & \hspace*{2.5em} \mbox{}
    + \Big| \sqrt{T_2 L_{2}}
        - \frac{\sqrt{T_2(1-L_{2})(1-T_2)L_{2}}\alpha^{\pm}(\Omega)}
            {1+\alpha^{\pm}(\Omega) \sqrt{(1-T_2)(1-L_{2})} } \Big|^2. 
\nonumber
\end{eqnarray}
It is immediately verified 
$S^{\pm}_{{\rm out}1}(\Omega)=S^{\pm}_{{\rm out}2}(\Omega)$ when 
the system is just the uncontrolled OPO described in Section II-B, 
i.e., $T_2=1$ and $L_2=0$. 
Also $L_2=1$ leads to $S^{\pm}_{{\rm out}2}(\Omega)=1~\forall\Omega$, 
implying that the CF loop loss will cause the overall degradation 
of squeezing level in frequency. 
The time delays appearing in $\alpha^\pm(\Omega)$ will affect on 
the control performance as well, particularly for the effective 
bandwidth in frequency. 
This will be seen later on. 
In what follows we assume that the CF loop is on resonance, i.e., 
$e^{i\omega_0(\tau_a+\tau_b)}=-1$.

\begin{figure}[h]
  \centering
  \includegraphics[width=0.789\linewidth]{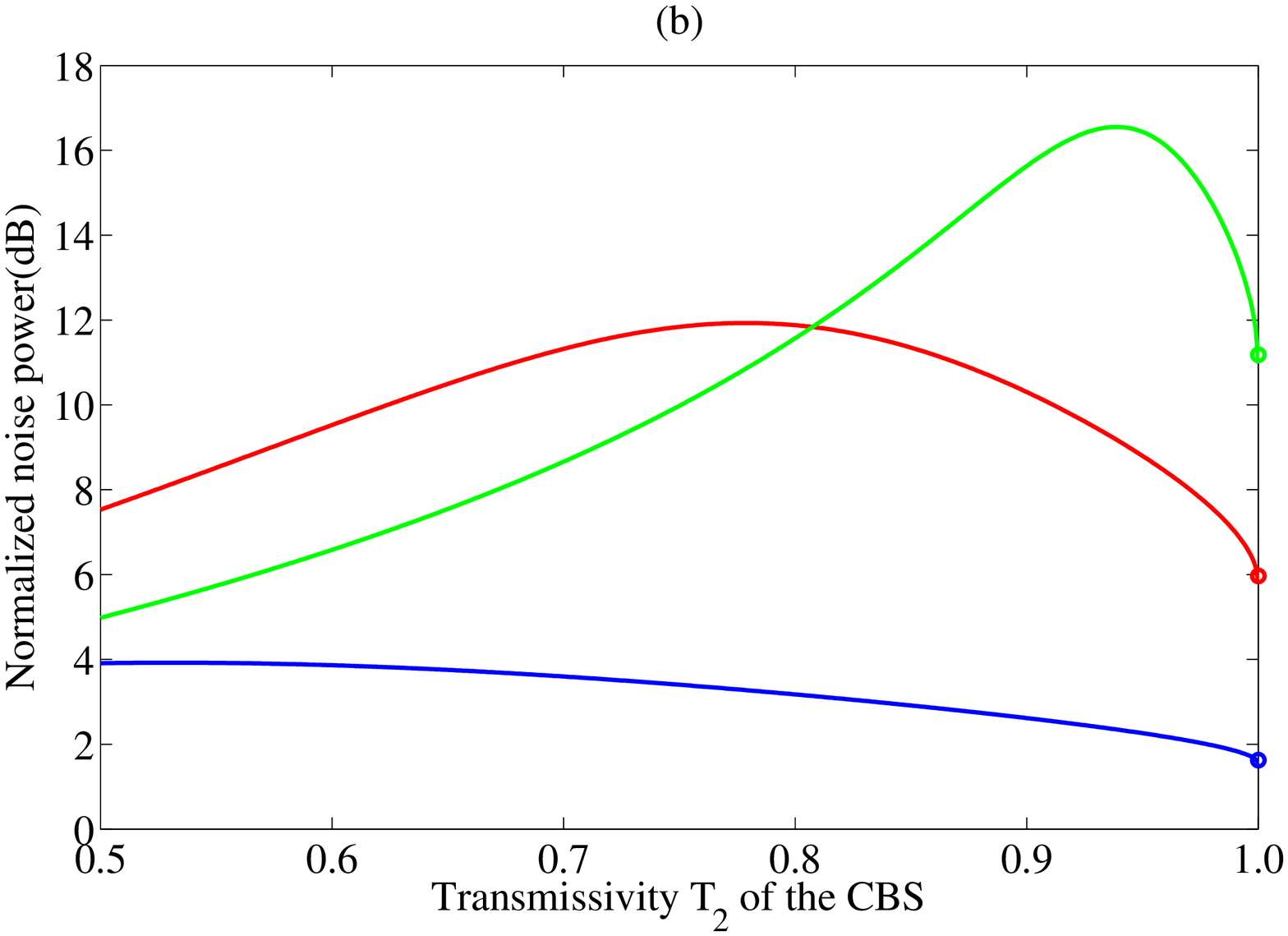}
  \includegraphics[width=0.8\linewidth]{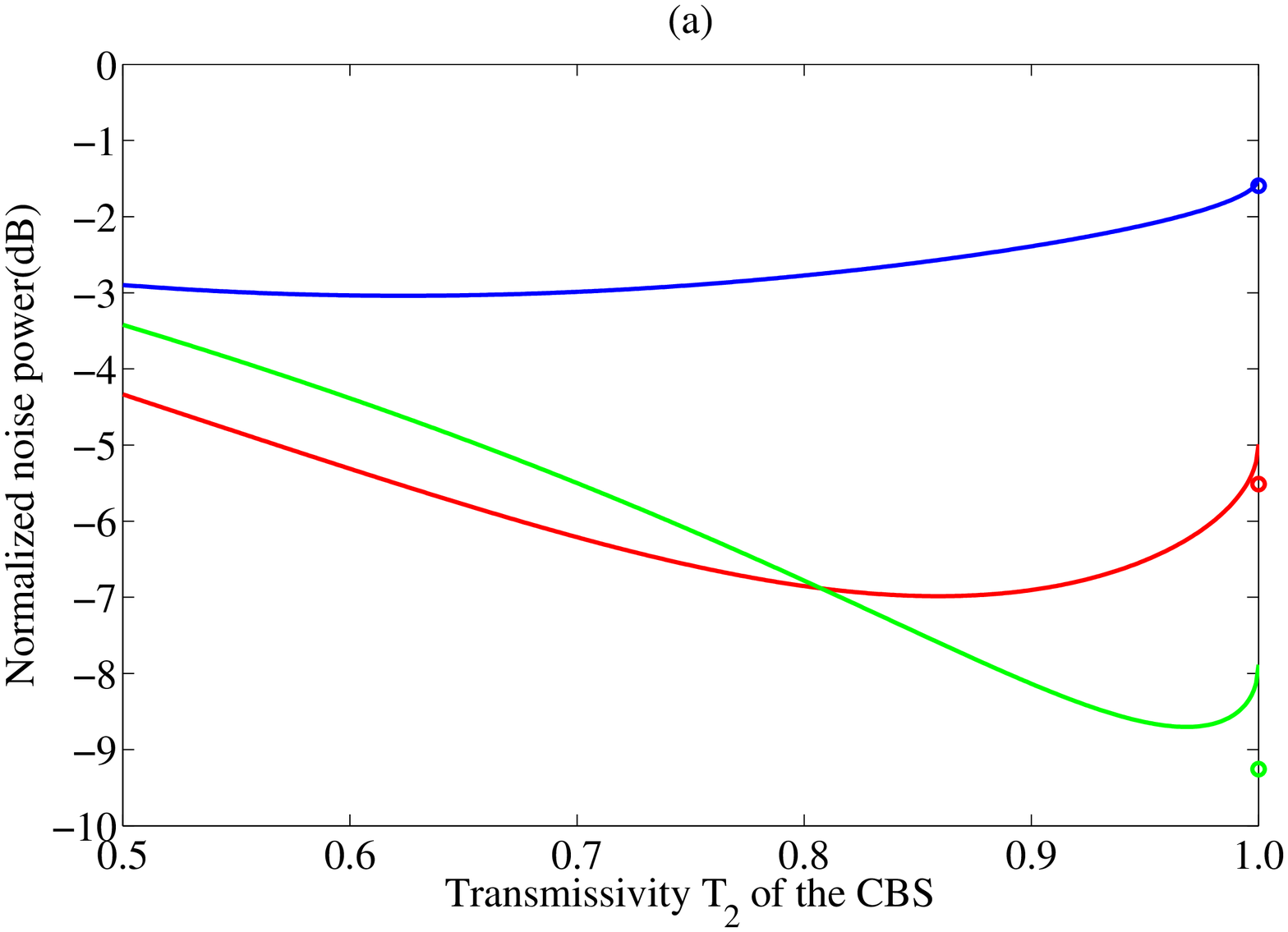}
  \caption{
  The transmissivity $T_2$ of the CBS versus the (a) anti-squeezing 
  and (b) squeezing levels for various normalized pumping 
  strength $x$. 
  The blue, red, and green lines correspond to $x=0.1$, 
  $x=0.35$, and $x=0.6$, respectively. 
  The circles indicate the values at $T_2=1$ with $L_2=0$, 
  corresponding to the uncontrolled OPO. 
  }
  \label{fig:T_BSdependence}
\end{figure}

Let us numerically evaluate the performance of how much the CF 
control can enhance the squeezing, or equivalently, can reduce 
the noise further. 
Here a set of practical values of parameters are taken 
\cite{Takeno-2007}: 
$T_1=0.12$, $L_{1}=5.0 \times 10^{-3}$, $L_{2}=5.0 \times 10^{-2}$, 
$l=0.5$~m, and $l_a=l_b=0.25$~m. 
To calculate $S^{\pm}_{{\rm out}2}(\Omega)$ we particularly focus 
on the values at frequency $\Omega/2\pi=1$~MHz. 
Fig. \ref{fig:T_BSdependence} depicts how the (a) anti-squeezing 
and (b) squeezing levels depend on $T_2$, with various values of 
the normalized pumping strength $x:=2|\epsilon|/\gamma$. 
Here, the power spectrum is shown in the unit of normalized 
magnitude, i.e., $10\log_{10}(S^{\pm}_{{\rm out}2}/S^{\pm}_{{\rm in}2})$ 
dB, where $S^{\pm}_{{\rm in}2}(\Omega)
:=\langle|\hat{X}^{\pm}_{{\rm in}2}(\Omega)|^2\rangle=1$ is the power 
of the vacuum input. 
Thus the horizontal axis ($0$~dB) corresponds to the QNL. 
Now the circles indicate the values at $T_2=1$ with $L_2=0$, i.e., 
the squeezing and anti-squeezing levels of the uncontrolled OPO. 
Then, in the case of weak pumping ($x=0.1$ or $x=0.35$), we find 
$T_2$ such that the squeezing level is enhanced by the CF control 
compared to that of the uncontrolled OPO. 
However, in the case of strong pumping ($x=0.6$), the CF control 
cannot enhance the squeezing at all. 
This is understood by considering the trade-off between the 
enhancement of the nonlinear squeezing effect and the CF loop loss; 
that is, the more strongly the CF control enhances the nonlinear effect, 
the more loss it must incur. 
Therefore, when the OPO is already pumped strongly, the CF loop 
loss becomes dominant compared to the enhancement of the nonlinear 
effect, and we cannot perform much enhancement of the squeezing. 
This is a limitation of the CF control for the squeezing 
enhancing problem. 
\begin{figure}[h]
  \centering
  \includegraphics[width=0.9\linewidth]{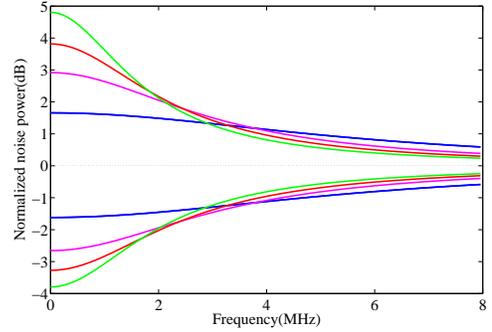}
  \caption{
  Frequency dependences of the squeezing and anti-squeezing levels. 
  The green, red, pink, and blue lines represent those under the 
  condition of $T_2$=0.7, 0.8, 0.9, and 1.0, respectively.}
  \label{fig:Frequency_dependence_theory}
\end{figure}

Next, to see the frequency-dependence of the CF control, 
we calculate $S^{\pm}_{{\rm out}2}(\Omega)$ with fixed value 
at $x=0.1$, which in the above discussion was proven to be a 
value such that the CF control has clear benefit. 
Fig. \ref{fig:Frequency_dependence_theory} shows the squeezing 
and anti-squeezing levels in the following cases: 
$T_2$=0.7, 0.8, 0.9, and 1.0. 
Now the squeezing and anti-squeezing levels of the uncontrolled 
OPO are almost the same as those with $T_2=1$ and $L_2=0.05$, which 
are indicated by the blue lines. 
Therefore, the squeezing enhancement can be evaluated by simply 
comparing the squeezing level with the CF ($T_2\neq 1$) to that 
without the CF ($T_2=1$), for a fixed value of $L_2$. 
(Note this argument makes sense only in the case of weak pumping 
power.) 
Then, in each case of $T_2$, the squeezing enhancement is observed 
only at lower frequencies. 
Moreover, while better squeezing is achieved by taking a smaller 
value of $T_2$, this brings the narrower effective bandwidth in 
frequency. 
This additional limiting property of the CF control is mainly 
due to the time delays occurred in the OPO and the feedback loop.


\section{The coherent feedback experiment}


\subsection{Experimental setup}

\begin{figure}[h]
  \centering
  \includegraphics[width=0.9\linewidth]{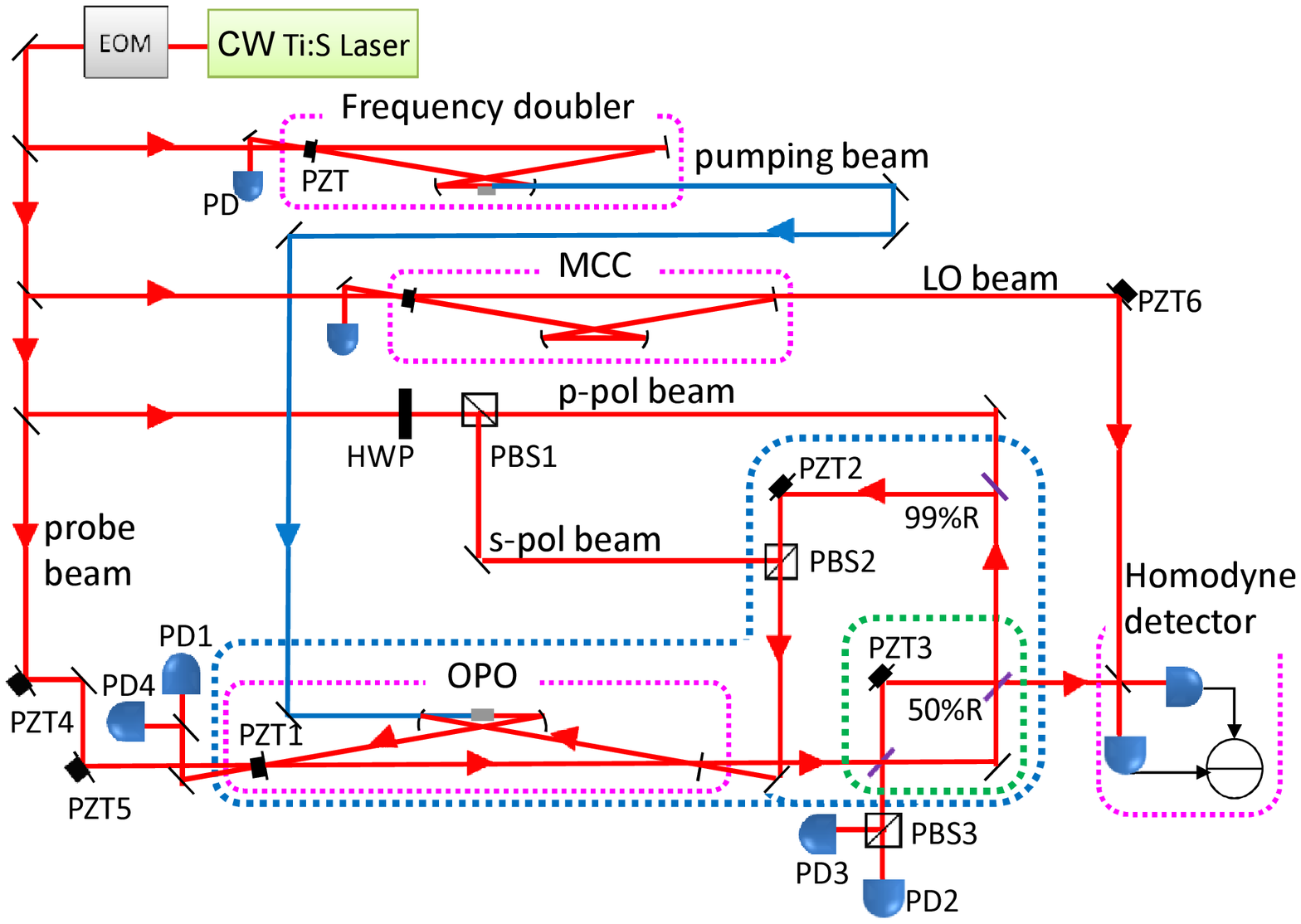}
  \caption{Experimental configuration. 
  OPO: optical parametric oscillator, 
  MCC: mode cleaning cavity, 
  PD: photo detector, 
  PZT: piezoelectric transducer, 
  PBS: polarized beam splitter, 
  HWP: half wave plate, and 
  LO: local oscillator. 
  The blue dashed line indicates the CF loop. 
  The green dashed line indicates the Mach-Zehnder interferometer, 
  which corresponds to the CBS. }
  \label{ExperimentalSetup}
\end{figure}

Fig. \ref{ExperimentalSetup} shows our experimental setup. 
The light source is a continuous-wave Ti:Sapphire laser 
(Coherent, MBR-110). 
The wavelength is 860 nm and the beam is horizontally polarized. 
A phase modulation of 10.4 MHz is applied on the beam for locking 
of all cavities by Pound-Drever-Hall method
\cite{Black2001,Boyd1996}.

The system is composed of four parts. 
The first is a frequency-doubler, which is a cavity to generate 
a second harmonic beam of 430 nm \cite{Masada2010}. 
This beam is used as a pumping beam for the OPO. 
The second is a mode-cleaning cavity that is used to clean up 
the spatial mode of local oscillator (LO) for homodyne detection 
so as to attain higher mode matching between the LO and 
the squeezed beam.

The third part consists of the OPO, the CBS, and the CF loop. 
The structure of the OPO is the same as in \cite{Takeno-2007}. 
In order to realize several locking, a {\it probe} beam is injected 
into the OPO from the high-reflection-coated mirror. 
The reflected beam is detected with PD1 and PD4 to get error 
signals for locking the cavity and the relative phase between the 
probe beam and the pumping beam. 
To lock the cavity we demodulate the output of the PD1 with 10.4 MHz 
modulation signal, and feed back the error signal to PZT1. 
On the other hand, to lock the relative phase between the probe 
beam and the pumping beam, we apply a phase modulation of 107 kHz 
on the probe beam with PZT4. 
We demodulate the output of the PD4 with 107 kHz modulation signal, 
and feed back the error signal to PZT5. 
Furthermore, we obtain the probe beam at the output port of the OPO 
which is used to lock the relative phase between the probe and 
the LO beams as explained later.

The CBS is realized by using a Mach-Zehnder (MZ) interferometer. 
The transmissivity can be determined by adjusting the phase 
difference between two arms in the MZ interferometer.
In order to lock a particular transmissivity, a s-polarized (s-pol) 
beam is injected into the CF loop from PBS2. 
Note that this s-pol beam does not circulate in the CF loop.
The beam is detected by PD3 to give the error signal of the CBS, 
which is fed back to PZT3. 
Additionally, to lock the CF loop, we inject a p-polarized (p-pol) 
beam into the CF loop from the mirror of 0.99 reflectivity. 
Note that this beam and the squeezed beam counter-propagate, hence 
this beam does not contaminate the CF output (the squeezed beam). 
We obtain the error signal by demodulating the output of PD2 with 
10.7 MHz modulation signal, and feed back it to PZT2.

The last part is homodyne detection. 
In order to measure a specific quadrature amplitude accurately, 
the relative phase between the probe beam (equivalently, the 
squeezed beam) and the LO beam should be locked. 
The error signal is obtained from the output of the homodyne 
detector by demodulating it with 107 kHz modulation signal. 
The error signal is fed back to PZT6. 
When measuring the squeezed beam, the probe beam is set to 
4 $\mu$W, and the LO beam is set to 3 mW, so that we can attain 
high signal-to-noise ratio without saturation of the homodyne 
detector. 
The output of the homodyne detector is measured with a spectrum 
analyzer.


\subsection{Results and discussion}

\begin{figure}[h]
  \centering
  \includegraphics[width=0.78\linewidth]{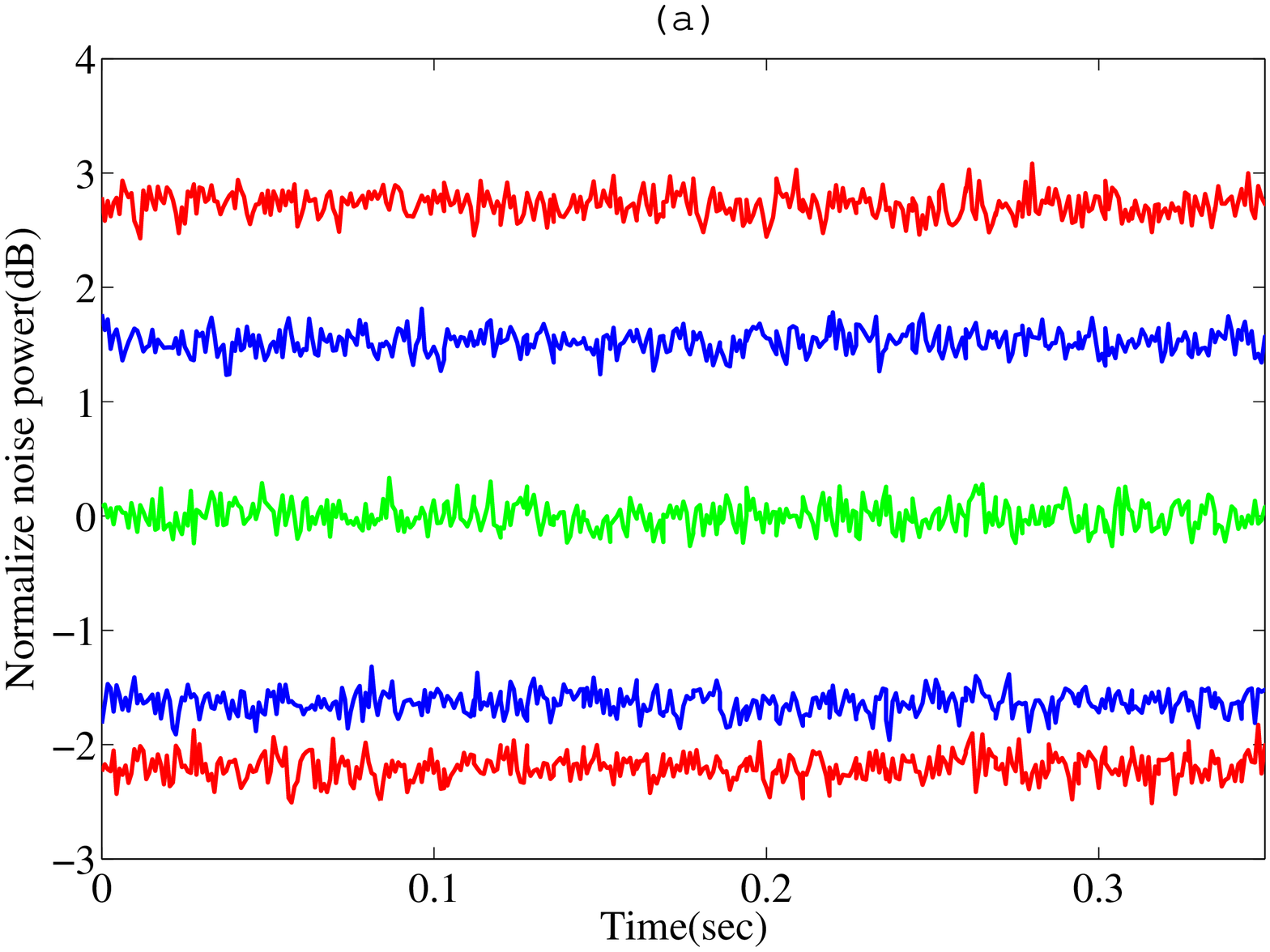}
  \includegraphics[width=0.78\linewidth]{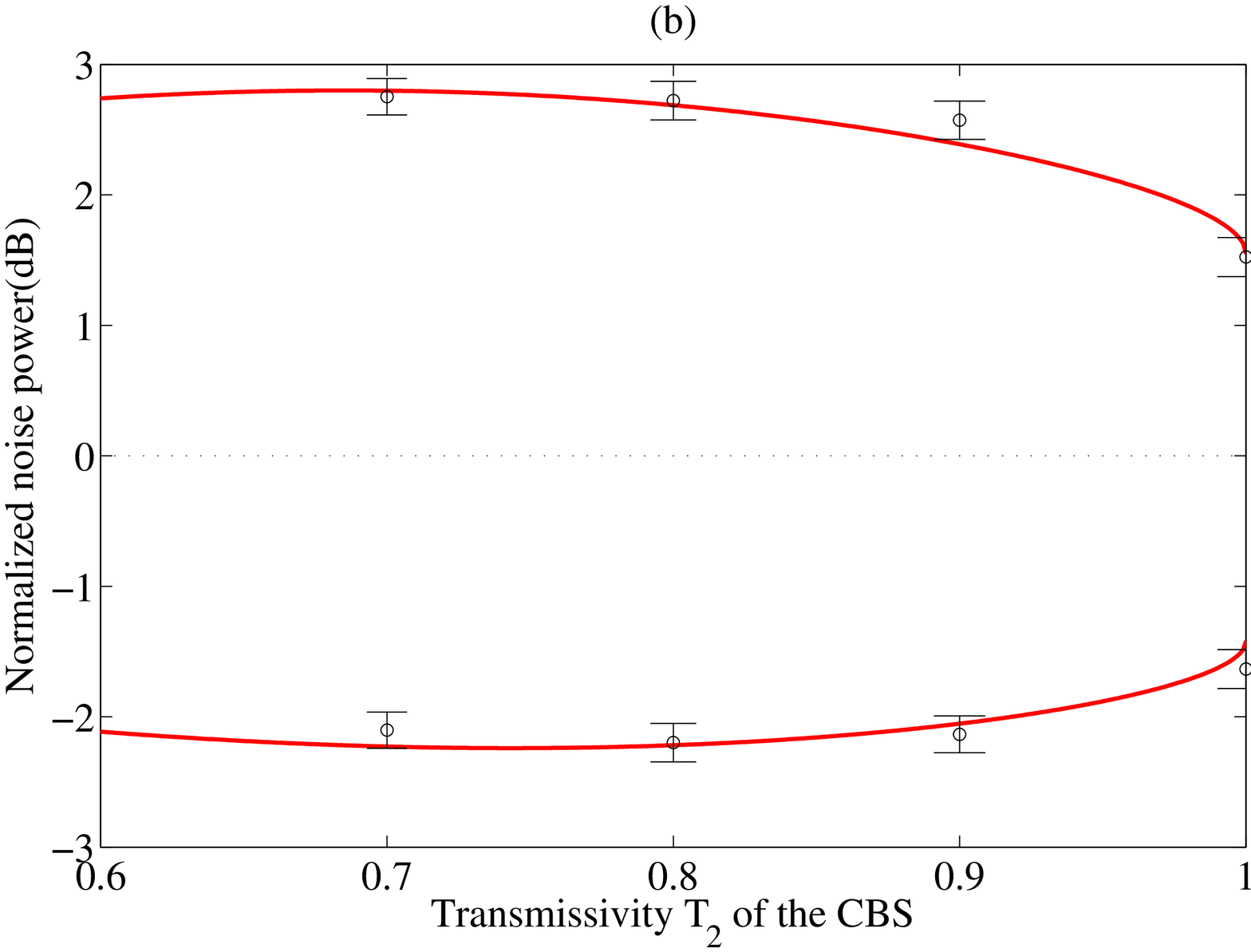}
  \caption{
  (a) Measurement results of the squeezing and anti-squeezing 
  levels at center frequency of 2.5 MHz. 
  The green line represents the vacuum noise level, i.e., the QNL. 
  The blue lines represent the squeezing and anti-squeezing levels 
  without the CF, and the red lines represent those with the CF when 
  $T_2=0.8$. 
  All the traces are averaged over 50 times. 
  Dark noise is subtracted. 
  (b) $T_2$-dependence of the squeezing and anti-squeezing 
  levels at center frequency of 2.5 MHz. 
  Circles and solid curves represent experimental and theoretical 
  values, respectively. 
  }
  \label{T-T2 dependence-exp}
\end{figure}

The parameters in this experiment are: 
$x=0.111$, $T_1=0.20$, $L_1=6.5 \times 10^{-3}$, $L_2=0.12$, 
$l=0.5$ m, and $l_a=l_b=0.25$ m. 
First we measure the squeezing and anti-squeezing levels with the 
CF ($T_2=0.8$) and those without the CF ($T_2=1.0)$. 
Note again that the squeezing enhancement can be evaluated by 
comparing these two values. 
Fig. \ref{T-T2 dependence-exp}~(a) shows the measurement results. 
The center frequency, the resolution bandwidth, and the video bandwidth
is $\Omega/2\pi=2.5$ MHz, 30 kHz, and 300 Hz, respectively. 
Here, because of a practical reason explained later, we cannot take 
the frequency $\Omega/2\pi=1$ MHz unlike the case discussed in 
Section III. 
The green line represents the vacuum noise level, i.e., the QNL. 
The blue lines represent the squeezing and anti-squeezing levels 
without the CF, and the red lines represent those with the CF. 
All the traces are normalized to the QNL. 
Fig. \ref{T-T2 dependence-exp} (a) clearly demonstrates the effect 
of the CF, showing the squeezing enhancement from 
$-$1.64$\pm$0.15 dB to $-$2.20$\pm$0.15 dB and the anti-squeezing 
enhancement from 1.52$\pm$0.15 dB to 2.72$\pm$0.15 dB.

We carry out measurements with several $T_2$; Fig. 
\ref{T-T2 dependence-exp} (b) shows $T_2$-dependence 
of the squeezing and anti-squeezing levels. 
Circles stand for the measurement results, and solid lines show the 
following theoretical values \cite{Takeno-2007}:
$S''_{\pm}(\Omega) = 1 + \eta ( S^{\pm}_{{\rm out}2}(\Omega)-1)$, 
where $\eta$ represents the overall detection efficiency given by 
$\eta = \xi^2  \rho$, $\xi$ is homodyne visibility and $\rho$ 
is quantum efficiency of photo diodes in the homodyne detector. 
In our experiment, we obtain $\eta=0.961$ with $\xi=0.985$, and 
$\rho=0.99$. 
Experimental and theoretical values show good agreement. 
Marginal gaps are attributed to fluctuation of the phase and the 
MZ interferometer locking.

\begin{figure}[h]
  \centering
  \includegraphics[width=0.9\linewidth]{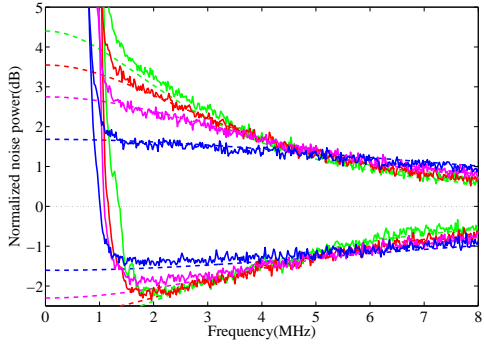}
  \caption{
  Frequency dependence of the squeezing and anti-squeezing levels. 
  The blue lines represent those without the CF, while the green, 
  red, and pink lines correspond to the case with the CF under the 
  condition of $T_2$=0.7, 0.8, and 0.9, respectively. 
  Dark noise is subtracted. 
  Dashed lines indicate theoretical values. }
  \label{fig:Frequency_dependence}
\end{figure}

A feature of the CF control can be seen in broadband measurement, 
where in this experiment we have observed that $x$ and $L_1$ are 
a bit changed to: $x=0.106$ and $L_{1}=9.0 \times 10^{-3}$.
Fig.~\ref{fig:Frequency_dependence} shows the frequency-dependence 
of the squeezing and anti-squeezing levels up to 8~MHz. 
The blue solid lines represent those without the CF, while the 
green, red, and pink solid lines correspond to the case with the 
CF under the condition of $T_2$=0.7, 0.8, and 0.9, respectively. 
At lower frequencies we find large laser noises and modulation 
signals used for several locking. 
This is the reason why, in our experiment, highly effective 
squeezing enhancement at for instance $\Omega/2\pi=1$ MHz cannot 
be observed; 
hence this is a practical limitation of the CF control. 
Except along such a noisy region in frequency, the results show 
good agreements with theoretical values illustrated with dashed 
lines, which are almost the same as those shown in 
Fig. \ref{fig:Frequency_dependence_theory}. 
That is, as discussed in Section III, better squeezing 
enhancement certainly brings the narrower effective bandwidth in 
frequency.



\section{Conclusion}

The first experimental demonstration of the CF control for 
squeezing enhancement is demonstrated. 
The results well agree with the theory that carefully takes 
into account the effects of the actual laboratory setup particularly 
time delays and losses occurring in the feedback loop. 
Although our feedback system is limited to linear optics, the 
results obtained in this work suggest realistic applicability 
of the CF control to various highly nonlinear quantum systems 
such as nanophotonic circuits \cite{Kerckhoff-2010,Mabuchi-2011}, 
which can be used for quantum error correction.



\end{document}